\documentstyle[amsmath,amssymb,epsfig,12pt]{article} 
 
\begin{document} 
\setlength{\parskip}{0.45cm} 
\setlength{\baselineskip}{0.75cm} 
%
%
%
\begin{titlepage} 
\setlength{\parskip}{0.25cm} 
\setlength{\baselineskip}{0.25cm} 
\begin{flushright} 
DO-TH 2002/07\\ 
\vspace{0.2cm} 
May 2002 
\end{flushright} 
\vspace{1.0cm} 
\begin{center} 
\LARGE 
{\bf The Polarized and Unpolarized Photon}\\ 
\LARGE{\bf Content of the Nucleon} 
\vspace{1.5cm} 
 
\large 
M. Gl\"uck, C.\ Pisano, E.\ Reya\\ 
\vspace{1.0cm} 
 
\normalsize 
{\it Universit\"{a}t Dortmund, Institut f\"{u}r Physik,}\\ 
{\it D-44221 Dortmund, Germany} \\ 
\vspace{0.5cm}

\vspace{1.5cm} 
\end{center} 
 
\begin{abstract} 
\noindent The equivalent photon content of polarized  
and unpolarized nucleons (protons, neutrons),  
utilized in Weizs\"acker--Williams approximations,  
are presented.  For this purpose a new expression  
for the elastic photon component of a polarized  
nucleon is derived. The inelastic photon components  
are obtained from the corresponding momentum  
evolution equations subject to the boundary  
conditions of their vanishing at some low momentum  
scale.  The resulting photon asymmetries, important  
for estimating cross section asymmetries in photon  
induced subprocesses are also presented for some  
typical relevant momentum scales. 
\end{abstract} 
\end{titlepage} 
 
 
 
The concept of the photon content of (charged)  
fermions is based on the equivalent photon  
(Weiz\"acker--Williams) approximation \cite{ref1}.   
Applied to the nucleon $N=p,\, n$ it consists of two  
parts, an elastic one due to $N\to \gamma N$ and  
an inelastic part due to $N\to\gamma X$ with  
$X\neq N$.  Accordingly the total photon distribution 
of the nucleon is given by  
\begin{equation} 
\gamma(y,Q^2) = \gamma_{e\ell}(y) + 
   \gamma_{ine\ell}(y,Q^2) 
\end{equation} 
where the elastic contribution of the proton,  
$\gamma_{e\ell}^p$, has been presented in \cite{ref2} 
which can be generally written as 
\begin{equation} 
\gamma_{e\ell}(y)=-\frac{\alpha}{2\pi}  
\int_{t_{\rm min}}^{t_{\rm max}} \frac{dt}{t} 
 \left\{ \left[2\left(\frac{1}{y}-1\right) 
  +\frac{2m^2y}{t}\right] H_1(t)+ yG_M^2(t)\right\} 
\end{equation} 
where $t\equiv q^2=-Q^2$ and 
\begin{equation} 
H_1(t)\equiv F_1^2(t)+\tau F_2^2(t) = 
  \frac{G_E^2(t) +\tau G_M^2(t)}{1+\tau} 
\end{equation} 
with $\tau\equiv -t/4m^2$, $m$ being the nucleon mass, 
and where $G_E=F_1-\tau F_2$ and $G_M=F_1+F_2$ are  
the common elastic (Sachs) form factors which are  
conveniently parametrized by the well known dipole  
form proportional to $(1-t/0.71$ GeV$^2)^{-2}$ as 
extracted from experiment.  For the proton, where 
$F_1^p(0)=1$ and $F_2^p(0)=\kappa_p\simeq 1.79$, we 
have 
\begin{equation} 
G_E^p(t)=(1+a\tau)^{-2}\,,\quad\quad 
 G_M^p(t)\simeq \mu_p G_E^p(t)\,,\quad\quad 
  H_1^p(t)=\frac{1+\mu_p^2\tau}{1+\tau}(1+a\tau)^{-4} 
\end{equation} 
with $\mu_p=1+\kappa_p\simeq 2.79$ and  
$a\equiv 4m^2/0.71$ GeV$^2\simeq 4.96$.  For the  
neutron, where $F_1^n(0)=0$ and  
$F_2^n(0)=\kappa_n\simeq -1.91$, we have 
\begin{equation} 
G_E^n(t) = \kappa_n\tau (1+a\tau)^{-2}\,, \quad\quad 
 G_M^n(t)=\kappa_n(1+a\tau)^{-2}\,, \quad\quad 
  H_1^n(t)=\kappa_n^2\tau(1+a\tau)^{-4}\, . 
\end{equation} 
In the relevant kinematic region $s \gg m^2$ the  
integration bounds in (2) can be approximated by 
$t_{\rm min}=-\infty$ and $t_{\rm max}=-m^2y^2/(1-y)$ 
so as to obtain an universal process independent  
$\gamma_{e\ell}(x)$.  Equation (2) can now be  
analytially integrated which yields for the proton 
\begin{equation} 
\gamma_{e\ell}^p(y) = \frac{\alpha}{2\pi}\, 
 \frac{2}{y} \left\{  
  \left[ 1-y +\frac{y^2}{4}(1+4a+\mu_p^2)\right]  
   I + (\mu_p^2-1)  
    \left[1-y+\frac{y^2}{4}\right] \tilde{I} - 
      \frac{1-y}{z^3} \right\} 
\end{equation} 
and for the neutron 
\begin{equation} 
\gamma_{e\ell}^n(y) = \frac{\alpha}{2\pi}\,\kappa_n^2 
 \frac{y}{2}  
  \left\{ I +\frac{1}{3}\,\frac{1}{(z-1)z^3}\right\} 
\end{equation} 
where $z\equiv 1+\frac{a}{4}\,\frac{y^2}{1-y}$ and 
\begin{eqnarray} 
I & = & \int_{\frac{y^2}{4(1-y)}}^{\infty} d\tau 
 \frac{1}{\tau(1+a\tau)^4} = -\ln \left(1-\frac{1}{z}\right) 
  - \frac{1}{z}-\frac{1}{2z^2} -\frac{1}{3z^3}\\ 
\tilde{I} & = & \int_{\frac{y^2}{4(1-y)}}^{\infty}d\tau 
 \frac{1}{(1+\tau)(1+a\tau)^4} = -\frac{1}{a_-^4}\ln 
  \left(1+\frac{a_-}{z}\right) +\frac{1}{a_-^3z} - 
   \frac{1}{2a_-^2z^2} +\frac{1}{3a_-z^3} 
\end{eqnarray} 
with $a_-=a-1$.  For arriving at (6) we have also 
utilized the relation 
\begin{displaymath} 
\int_{\frac{y^2}{4(1-y)}}^{\infty} d\tau 
 \frac{1}{\tau^2(1+a\tau)^4} = -4aI +4\frac{1-y}{y^2z^3} 
\end{displaymath} 
which will be also relevant for the polarized photon 
contents to be presented below.  Our result in (6) agrees 
with the one presented in a somewhat different form in 
\cite{ref2}.  Finally, the inelastic part in (1) has  
been given in \cite{ref3}, 
\begin{equation} 
\frac{d\gamma_{ine\ell}^N(y,Q^2)}{d\ln Q^2} = 
 \frac{\alpha}{2\pi} \sum_{q=u,d,s} e_q^2 
  \int_y^1 \frac{dx}{x}\, P_{\gamma q} 
   \left(\frac{y}{x}\right)  
    \left[ q^N(x,Q^2)+\bar{q}^N(x,Q^2)\right] 
\end{equation} 
with $P_{\gamma q}(x) = [1+(1-x)^2]/x$ and where 
$\stackrel{\!\!\!(-)}{q^p}\equiv\stackrel{(-)}{q}$ and 
$\stackrel{\!\!\!(-)}{u^n} = \stackrel{\,(-)}{d}$, 
$\stackrel{\!\!\!(-)}{d^n} = u$,  
$\stackrel{\!\!\!(-)}{s^n} = 
\stackrel{(-)}{s}$.  This equation was integrated 
subject to the  `minimal' boundary condition 
$\gamma_{ine\ell}^N(y,Q_0^2)=0$ at \cite{ref4} 
$Q_0^2=0.26$ GeV$^2$, which is obviously not  
compelling and affords further theoretical and  
experimental studies.  Since for the time being 
there are no experimental measurements available, 
the  `minimal' boundary condition provides at  
present a rough estimate for the inelastic 
component at $Q^2\gg Q_0^2$. 
 
Clearly, the nucleon's photon content $\gamma^N 
(x,Q^2)$ is not such a fundamental quantity as are  
its underlying parton distributions $f(x,Q^2)=q,\, 
\bar{q},\, g$ or the parton distributions $f^{\gamma} 
(x,Q^2)$ of the photon, since $\gamma^p(x,Q^2)$  
is being derived from these more fundamental 
quantities.  It represents mainly a technical device 
which allows for a simpler and more efficient 
calculation of photon--induced subprocesses. For 
example, the analysis of the deep inelastic Compton 
scattering process $ep\to e\gamma X$ reduces  
\cite{ref3,ref5} to the calculation of the $2\to 2$ 
subprocess $e\gamma\to e\gamma$ instead of having 
to calculate the full $2\to 3$ subprocess 
$eq\to e\gamma q$.  Similar remarks 
hold for the production of charged heavy particles 
(e.g.\ Higgses) via $\gamma\gamma$ fusion at high 
energy $pp$ colliders, $pp\to\gamma\gamma X\to 
H^+H^-X$.  The reliability of this approximation 
remains, however, to be studied.  
 
Our main purpose here is to extend these calculations 
to the polarized sector, i.e., to 
\begin{equation} 
 \Delta\gamma(y,Q^2)=\Delta\gamma_{e\ell}(y)+ 
  \Delta\gamma_{ine\ell}(y,Q^2)\, . 
\end{equation} 
The inelastic contribution derives from a  
straightforward extension of eq.\, (10), 
\begin{equation} 
\frac{d\Delta\gamma_{ine\ell}^N(y,Q^2)}{d\ln Q^2} 
 = \frac{\alpha}{2\pi} \sum_{q=u,d,s} e_q^2 
  \int_y^1\, \frac{dx}{x}\, \Delta P_{\gamma q} 
   \left( \frac{y}{x}\right)  
    \left[\Delta q^N(x,Q^2)+\Delta\bar{q}^N(x,Q^2) 
      \right] 
\end{equation} 
where $\Delta P_{\gamma q}(x)=[1-(1-x)^2]/x=2-x$. 
We integrate this evolution equation assuming again 
the not necessarily compelling  `minimal' boundary 
condition $\Delta\gamma_{ine\ell}^N(y,Q_0^2)=0$, 
according to $|\Delta\gamma_{ine\ell}(y,Q_0^2)|\leq 
\gamma_{ine\ell}(y,Q_0^2)=0$, at $Q_0^2=0.26$ GeV$^2$ 
using the recent LO polarized parton densities of 
\cite{ref6}.  
 
The elastic distribution $\Delta\gamma_{e\ell}(y)$ in 
(11) is determined via the antisymmetric part of the 
tensor describing the photon emitting fermion  
(nucleon) 
\begin{equation} 
T^{\mu\nu}=Tr 
 \left[ \frac{1}{2}(1+\gamma_5n\!\!\!/) 
  (p\!\!\!/+m)\Gamma^{\mu}(p\!\!\!/\,'+m)\Gamma^{\nu} 
   \right] 
\end{equation} 
for the generic process 
\begin{equation} 
  N(p;\, n)+a(k;\, s)\to  N({p'}) + X 
\end{equation} 
where $a$ being a suitable target (parton, photon, 
etc.) with momentum $k$ and $n$, $s$ are the  
appropriate polarization vectors \cite{ref7}  
satisfying $n\cdot p=0$ and $s\cdot k=0$.  In terms 
of the Dirac and Pauli form factors $F_{1,2}(t)$ of 
the nucleon the elastic vertices $\Gamma^{\mu}$ are 
given by 
\begin{equation} 
\Gamma^{\mu}=(F_1+F_2)\gamma^{\mu}-\frac{1}{2m} 
  F_2(p+{p'})^{\mu}\,. 
\end{equation} 
The analysis is now a straightforward extension of 
the calculation \cite{ref7} of the polarized 
Weizs\"acker--Williams distribution resulting from 
a photon emitting fermion (electron) where $N\to e$ 
in (14) with $\Gamma^{\mu}=\gamma^{\mu}$, and all 
relevant definitions and kinematics can be found 
in \cite{ref7} as well.  The resulting antisymmetric 
part 
\footnote[1]{It should be noted that the symmetric 
(unpolarized) tensor 
\begin{eqnarray}  
T_S^{\mu\nu} & = & Tr\left[ (p\!\!\!/+m)\Gamma^{\mu} 
     (p\!\!\!/\,'+m)\Gamma^{\nu}\right]\nonumber\\ 
 & = & 4 G_M^2 \left[p^{\mu}{p'}^{\nu}+{p'}^{\mu}p^{\nu} 
     +\frac{q^2}{2} \, g^{\mu\nu} \right]  
       -4(p+{p'})^{\mu} (p+{p'})^{\nu}  
        \left[G_M F_2 - \frac{1}{2} 
          \left( 1-\frac{q^2}{4m^2}\right)F_2^2  
            \right]\nonumber 
\end{eqnarray} 
gives rise to the same Weizs\"acker--Williams  
distribution obtained in a somewhat less transparent 
way in [2], i.e.\ to eq.\ (2), when the analysis 
\protect\cite{ref8} for a photon emitting pointlike unpolarized 
fermion (electron) is straightforwardly extended to 
an unpolarized nucleon, $N(p)+a(k)\to N({p'})+X$, 
instead to the polarized process (14).}  
of $T^{\mu\nu}$ is 
\begin{equation} 
T_A^{\mu\nu}= 2im\,G_M^2\,\varepsilon^{\mu\nu\rho\sigma} 
  n_{\rho}q_{\sigma}+2i\, G_M(F_2/2m)\left[(p+{p'})^{\mu}\, 
   \varepsilon^{\nu\rho\sigma\sigma'} -(p+{p'})^{\nu} 
    \varepsilon^{\mu\rho\sigma\sigma'}\right] 
     n_{\rho}p_{\sigma}{p'}_{\sigma'} 
\end{equation} 
with $q=p-{p'}$.  It is now straightforward to contract 
$T_A^{\mu\nu}$ with the appropriate antisymmetric part 
of the tensor $W_A^{\mu\nu}$ describing the polarized 
target $a(k;\,s)$ in (14) which is expressed in terms 
of the usual polarized structure functions $g_1$ and 
$g_2$ where all terms proportional to $g_2$ drop in 
$T_A\cdot W_A$.  This yields 
\begin{eqnarray} 
\Delta\gamma_{e\ell}(y) & = & -\frac{\alpha}{2\pi} 
   \int_{t_{\rm min}}^{t_{\rm max}} \frac{dt}{t} 
    \left\{ \left[2-y + \frac{2m^2y^2}{t}\right] 
     G_M^2(t)-2\left[1-y + \frac{m^2y^2}{t}\right] 
      G_M(t)F_2(t)\right\}\nonumber\\ 
& = & -\frac{\alpha}{2\pi}  
     \int_{t_{\rm min}}^{t_{\rm max}} \frac{dt}{t}\, 
      G_M(t)\left\{ \left[2-y +\frac{2m^2y^2}{t}\right] 
       F_1(t) + yF_2(t)\right\} 
\end{eqnarray} 
with $y=k\cdot q\,/\, k\cdot p$ and the first term  
proportional to $G_M^2$ in the first line corresponds 
to the pointlike result of \cite{ref7}. Following 
\cite{ref2}, we again approximate the integration 
bounds by $t_{\rm min}=-\infty$ and $t_{\rm max}= 
-m^2y^2/(1-y)$ as in (2) in order to obtain an  
universal process independent polarized elastic  
distribution.  Using, in addition to (4) and (5), 
\begin{eqnarray} 
F_1^p(t) & = & \frac{1+\mu_p\tau}{1+\tau}\, 
               (1+a\tau)^{-2},\quad 
      F_2^p(t) = \frac{\kappa_p}{1+\tau} 
                  (1+a\tau)^{-2}\\ 
F_1^n(t) & = & 2\kappa_n \frac{\tau}{1+\tau} 
               (1+a\tau)^{-2},\quad 
      F_2^n(t) = \kappa_n \frac{1-\tau}{1+\tau} 
                  (1+a\tau)^{-2}\, , 
\end{eqnarray} 
eq.\ (17) yields for the proton 
\begin{equation} 
\Delta\gamma_{e\ell}^p(y) = \frac{\alpha}{2\pi}\,\mu_p 
 \left\{ \left[ (2-y) 
  \left( 1+\kappa_p\frac{y}{2} \right)  
   +2ay^2\right] I +2\kappa_p(1-y+\frac{y^2}{4}) 
    \tilde{I} -2\, \frac{1-y}{z^3}\right\}\, , 
\end{equation} 
and for the neutron  
\begin{equation} 
\Delta\gamma_{e\ell}^n(y) = \frac{\alpha}{2\pi}  
 \kappa_n^2 \left\{ y(1-y) I +4 ( 1-y + 
  \frac{y^2}{4} )\tilde{I}\right\} 
\end{equation} 
with $I$ and $\tilde{I}$ being given in (8) and (9). 
These latter two equations together with (12) for 
$N=p,\, n$ yield now the total photon content  
$\Delta\gamma^N(y,Q^2)$ of a polarized nucleon in (11). 
 
Our results for $\Delta\gamma^p(y,Q^2)$ in (11) are 
shown in fig.\ 1 for some typical values of $Q^2$ up 
to $Q^2=M_W^2=6467$ GeV$^2$.  For comparison the 
expectations for the unpolarized $\gamma^p(y,Q^2)$ 
in (1) are depicted as well.  The $Q^2$--independent 
polarized and unpolarized elastic contribtions in 
eq.\ (20) and (6), respectively, are also shown 
separately. Due to the singular small--$x$ behavior 
of the unpolarized parton distributions  
$x\!\!\stackrel{(-)}{q}\!\!(x,Q^2)$ in (10) as well as of 
the singular $y\gamma_{e\ell}^p(y)$ in (6) as  
$y\to 0$, the total $y\gamma^p(y,Q^2)$ in fig.\ 1  
increases as $y\to 0$, whereas the polarized  
$y\Delta\gamma^p(y,Q^2)\to 0$ as $y\to 0$ because of  
the vanishing of the polarized parton distributions 
$x\Delta\!\!\stackrel{(-)}{q}\!\!(x,Q^2)$ in (12) 
at small $x$ and of the vanishing  
$y\Delta\gamma_{e\ell}^p(y)$ in (20) at small $y$. 
In fact, $y\Delta\gamma^p(y,Q^2)$ is negligibly 
small for $y$  
\raisebox{-0.1cm}{$\stackrel{<}{\sim}$} $10^{-3}$ 
as compared to $y\gamma^p(y,Q^2)$.  For larger 
values of $y$, $y>10^{-2}$, $y\Delta\gamma^p(y,Q^2)$  
becomes sizeable and in particular is dominated by  
the $Q^2$--independent elastic contribution  
$y\Delta\gamma_{e\ell}^p(y)$ at moderate values of 
$Q^2$, $Q^2$ 
\raisebox{-0.1cm}{$\stackrel{<}{\sim}$} 100 GeV$^2$  
(with a similar behavior in the unpolarized sector).   
This is evident from fig.\ 2 where the  
results of fig.\ 1 are plotted versus a linear 
$y$ scale.  The asymmetry $A_{\gamma}^p(y,Q^2)$ 
is shown in fig.\ 3 where 
\begin{equation} 
A_{\gamma}(y,Q^2)\equiv  
  \left[ \Delta\gamma_{e\ell}(y) 
   +\Delta\gamma_{ine\ell} (y,Q^2)\right] 
     /\gamma(y,Q^2) 
\end{equation} 
with the total unpolarized photon content of the  
nucleon being given by (1).  To illustrate the size 
of $\Delta\gamma_{e\ell}^p$ relative to the  
unpolarized $\gamma_{e\ell}^p$, we also show the 
$Q^2$--independent ratio  
$\Delta\gamma_{e\ell}^p(y)/\gamma_{e\ell}^p(y)$ 
in fig.\ 3 which approaches 1 as $y\to 1$. 
 
The polarized photon distributions $\Delta\gamma^p 
(y,Q^2)$ shown thus far always refer to the  
so called  `valence' scenario \cite{ref6} where the 
polarized parton distributions in (12) have  
flavor--broken light sea components $\Delta\bar{u} 
\neq \Delta\bar{d}\neq\Delta\bar{s}$, as is the  
case (as well as experimentally required) for the 
unpolarized ones in (10) where $\bar{u}\neq\bar{d} 
\neq\bar{s}$.  Using instead the somehow unrealistic 
`standard' scenario \cite{ref6} for the polarized 
parton distributions with a flavor--unbroken sea 
component $\Delta\bar{u}=\Delta\bar{d}=\Delta\bar{s}$, 
all results shown in figs.\ $1-3$ remain practically 
almost undistinguishable.  The same holds true for 
the photon content of a polarized neutron to which 
we now turn. 
 
The results for $\Delta\gamma^n(y,Q^2)$ are  
shown in fig.\ 4 which are sizeably smaller than 
the ones for the photon in fig.\ 1 and, furthermore, 
the elastic contribution is dominant while the  
inelastic ones become marginal at  
$y$ \raisebox{-0.1cm}{$\stackrel{>}{\sim}$} 0.2. 
For comparison the unpolarized $\gamma^n(y,Q^2)$ 
in (1) is shown in fig.\ 4 as well.  Here, 
$\gamma_{e\ell}^n$ in (7) is marginal and  
$y\gamma_{e\ell}^n(y)$ is non--singular as $y\to 0$ 
with a limiting value $y\gamma_{e\ell}^n(y)/\alpha 
=\kappa_n^2/(3\pi a)\simeq 0.078$.  Thus the  
increase of $y\gamma^n(y,Q^2)$ at small $y$ is  
entirely caused by inelastic component 
$y\gamma_{ine\ell}^n(y,Q^2)$ in (10), due to the 
singular small--$x$ behavior of  
$x\!\!\stackrel{(-)}{q}\!\!(x,Q^2)$, which is in  
contrast to $y\gamma^p(y,Q^2)$ in fig.~1.  These 
facts are more clearly displayed in fig.\ 5 where 
the results of fig.\ 4 are presented for a linear 
$y$ scale.  Notice that again the polarized 
$y\Delta\gamma^n(y,Q^2)\to 0$ as $y\to 0$ because 
of the vanishing of the polarized parton  
distributions  
$x\Delta\!\!\stackrel{(-)}{q}\!\!(x,Q^2)$ in (12)  
at small $x$ and of the vanishing of 
$y\Delta\gamma_{e\ell}^n(y)$ in (21) at small $y$. 
Finally, the asymmetry $A_{\gamma}^n(y,Q^2)$ 
defined in (22) is shown in fig.\ 6 which is  
entirely dominated by the elastic contribution for 
$x$ \raisebox{-0.1cm}{$\stackrel{>}{\sim}$} 0.2. 
As in fig.\ 3 we illustrate the size of the elastic 
$\Delta\gamma_{e\ell}^n(y)$ relative to the 
unpolarized $\gamma_{e\ell}^n(y)$ by showing the 
ratio $\Delta\gamma_{e\ell}^n/\gamma_{e\ell}^n$ 
in fig.\ 6 as well.  Notice that  
$\Delta\gamma_{e\ell}^n/\gamma_{e\ell}^n\to\frac 
{6}{7}$ as $y\to 1$ in contrast to the case of  
the proton.  
 
As mentioned at the beginning the knowledge of the  
unpolarized photon content of the nucleon  
$\gamma^N(y,Q^2)$ allows for a simpler and more 
efficient calculation of photon--induced subprocesses 
in elastic and deep inelastic $ep$ and purely  
hadronic ($pp,\, \ldots$) reactions  
\cite{ref2,ref3,ref5,ref9,ref10,ref11,ref12}. 
For example, to consider just the simple $2\to 2$ 
subprocess \mbox{$e\gamma\to e\gamma$}  
for the analysis of  
the deep inelastic Compton process $ep\to e\gamma X$ 
or $e\gamma\to\nu W$ for associated $\nu W$ production 
in $ep\to\nu WX$.  Similarly, the $\gamma\gamma$ 
fusion process $\gamma\gamma\to\ell^+\ell^-,\,  
c\bar{c},\, H^+H^-,\, \tilde{\ell}^+\tilde{\ell}^-,\, 
\ldots$ for (heavy) lepton ($\ell$), heavy quark  
($c$), charged Higgs ($H^{\pm}$) and slepton 
($\tilde{\ell}$) production etc.\ can be easily  
analyzed in purely hadronic $pp$  
reactions which is also an interesting possibility 
of producing charged particles which do not have  
strong interactions.  In particular the 
$\gamma\gamma\to\mu^+\mu^-$ channel will give 
access to experimental measurements of  
$\gamma^N(y,Q^2=M_{\mu^+\mu^-}^2)$ at $pp,\, pd$  
and $dd$ colliders. 
 
Analogous remarks hold for the longitudinally  
polarized $\vec{e}\vec{N}$ and $\vec{p}\vec{p},\, 
\vec{p}\vec{d}$ and $\vec{d}\vec{d}$ reactions 
where the polarized photon content of the nucleon 
$\Delta\gamma^N(y,Q^2)$, as calculated and studied 
in this article, enters.  Very interestingly, it 
remains to be seen whether ongoing experiments at 
RHIC(BNL) for dimuon production, $\vec{p}\vec{p},\, 
\vec{d}\vec{d}\to\mu^+\mu^-X$, can directly  
delineate and test our predictions for  
$\Delta\gamma^N(y,M_{\mu^+\mu^-}^2)$.  
   
A FORTRAN package (grids) containing our results 
for $\Delta\gamma^N(y,Q^2)$ as well as those for 
$\gamma^N(y,Q^2)$ can be obtained by electronic 
mail. 
\vspace{1.5cm} 
 
This work has been supported in part by the 
 `Bundesministerium f\"ur Bildung und Forschung', 
Berlin/Bonn. 
 
\newpage 
 
\noindent{\large{\bf{\underline{Figure Captions}}}} 
\begin{itemize} 
\item[\bf{Fig.\ 1}.]  The polarized and unpolarized 
      total photon contents of the proton,  
      $\Delta\gamma^p$ and $\gamma^p$, according to 
      eqs.\ (1) and (11) at some typical fixed values 
      of $Q^2$ (in GeV$^2$).  The $Q^2$--independent 
      elastic contributions are given by eqs.\ (20) 
      and (6). 
 
\item[\bf{Fig.\ 2}.]  As in fig.\ 1 but for a linear 
      $y$ scale. 
 
\item[\bf{Fig.\ 3}.]  The asymmetry of the polarized 
      to the unpolarized photon content of the proton 
      as defined in (22) at various fixed values of  
      $Q^2$ (in GeV$^2$) according to the results in 
      fig.\ 1.  The $Q^2$--dependence of the elastic 
      contribution to $A_{\gamma}^p$ is caused by 
      the $Q^2$--dependent total unpolarized photon 
      content in the denominator of (22).  For 
      illustration the $Q^2$--independent elastic ratio 
      $\Delta\gamma_{e\ell}^p/\gamma_{e\ell}^p$ is 
      shown as well. 
 
\item[\bf{Fig.\ 4}.]  As fig.\ 1 but for the neutron, 
      with elastic polarized and unpolarized contributions 
      being given by eqs.\ (21) and (7). 
 
\item[\bf{Fig.\ 5}.]  As in fig.\ 4 but for a linear 
      $y$ scale. 
 
\item[\bf{Fig.\ 6}.] As fig.\ 3 but for the neutron 
      asymmetry according to the results in fig.\ 4. 
\end{itemize} 
 
\newpage 
\pagestyle{empty} 
\begin{figure} 
\centering 
\vspace*{-1cm} 
\hspace*{-1.5cm} 
\epsfig{figure=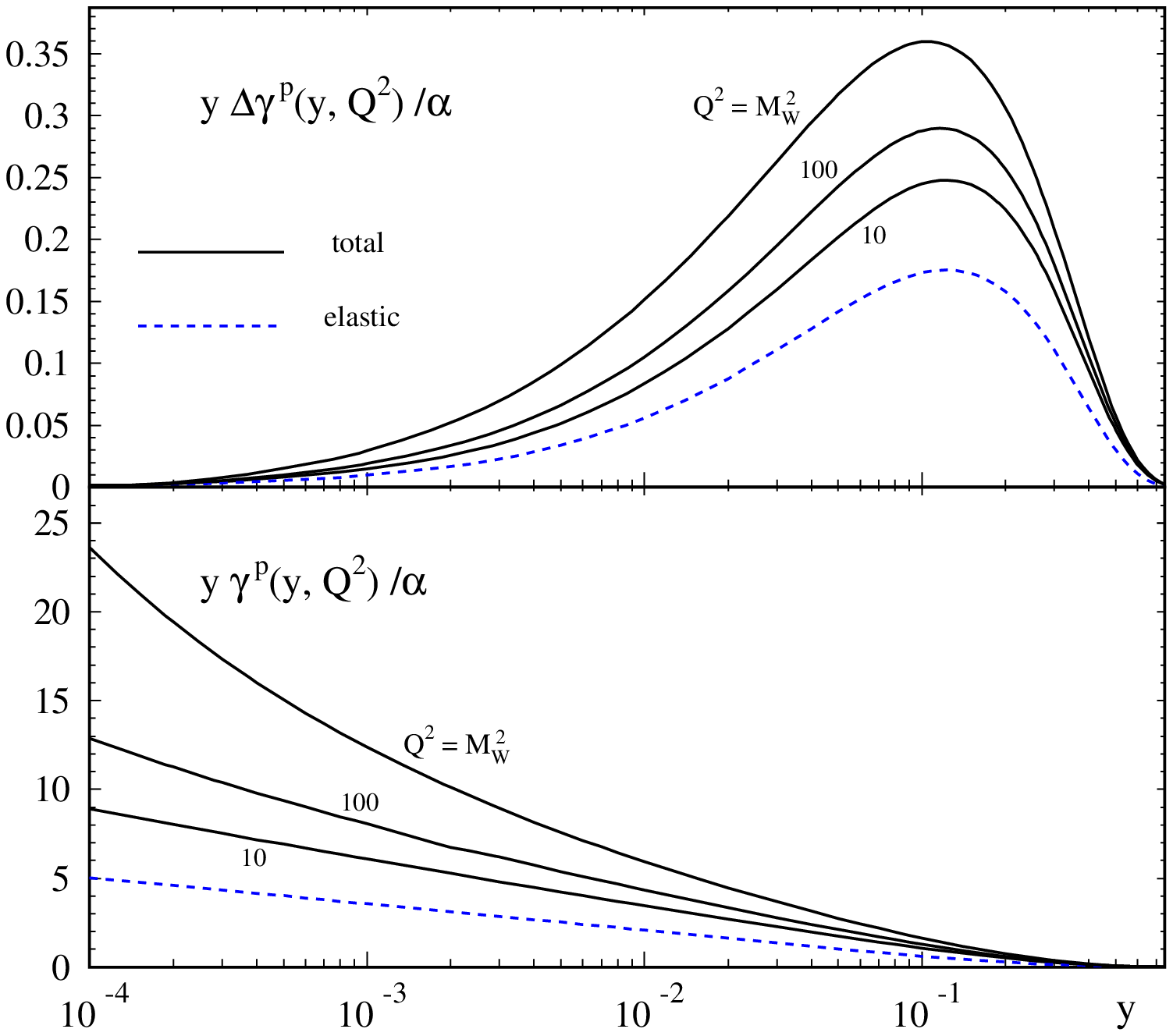,width=18cm} 
 
\vspace*{2.cm} 
{\large\bf Fig. 1} 
\end{figure} 
 
\newpage 
\begin{figure}[t] 
\centering 
\hspace*{-1.5cm} 
\epsfig{figure=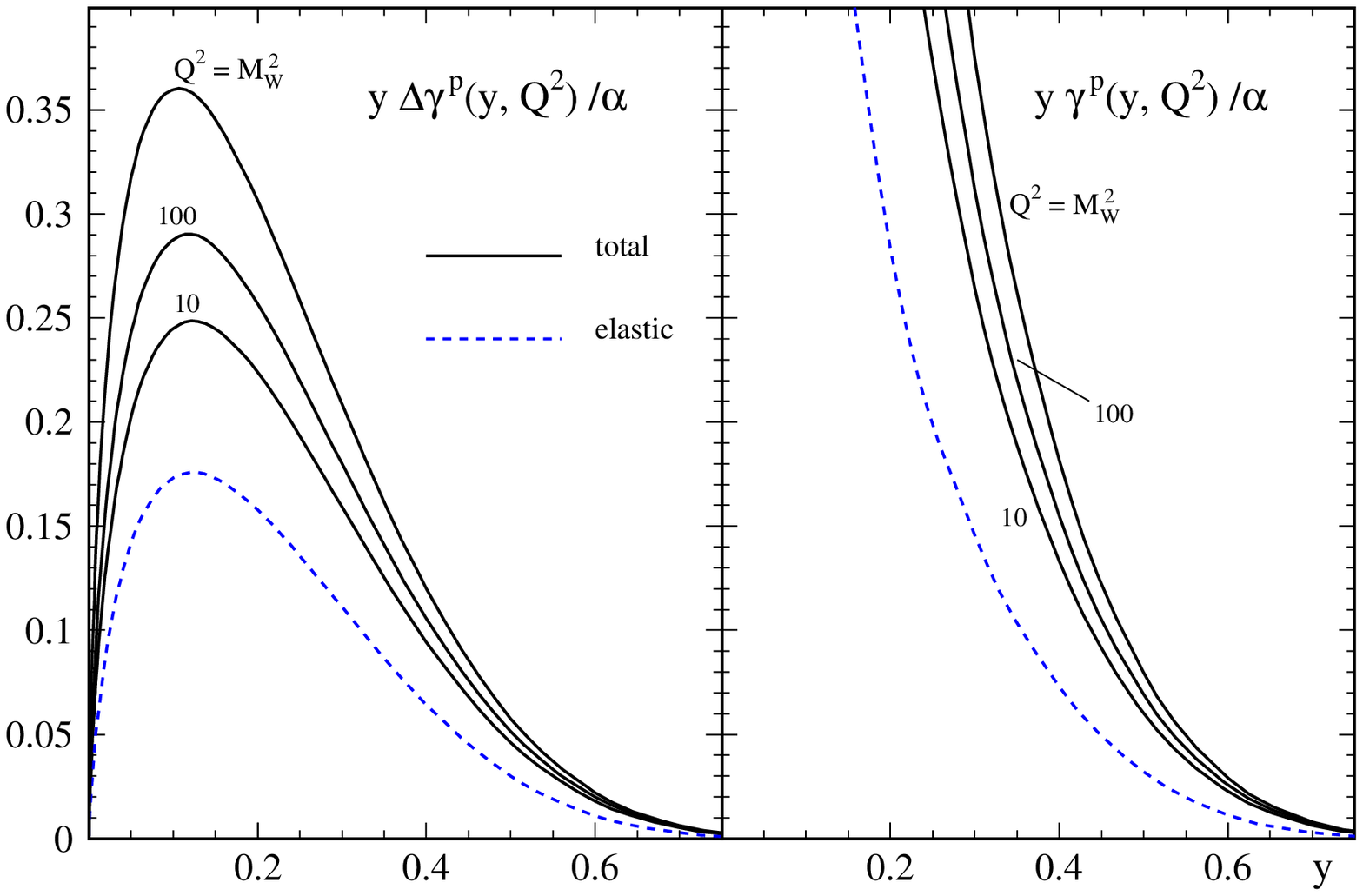,width=18cm} 
 
\vspace*{2cm} 
{\large\bf Fig. 2} 
\end{figure} 
 
\newpage 
\begin{figure}[t] 
\centering 
\hspace*{-1.5cm} 
\epsfig{figure=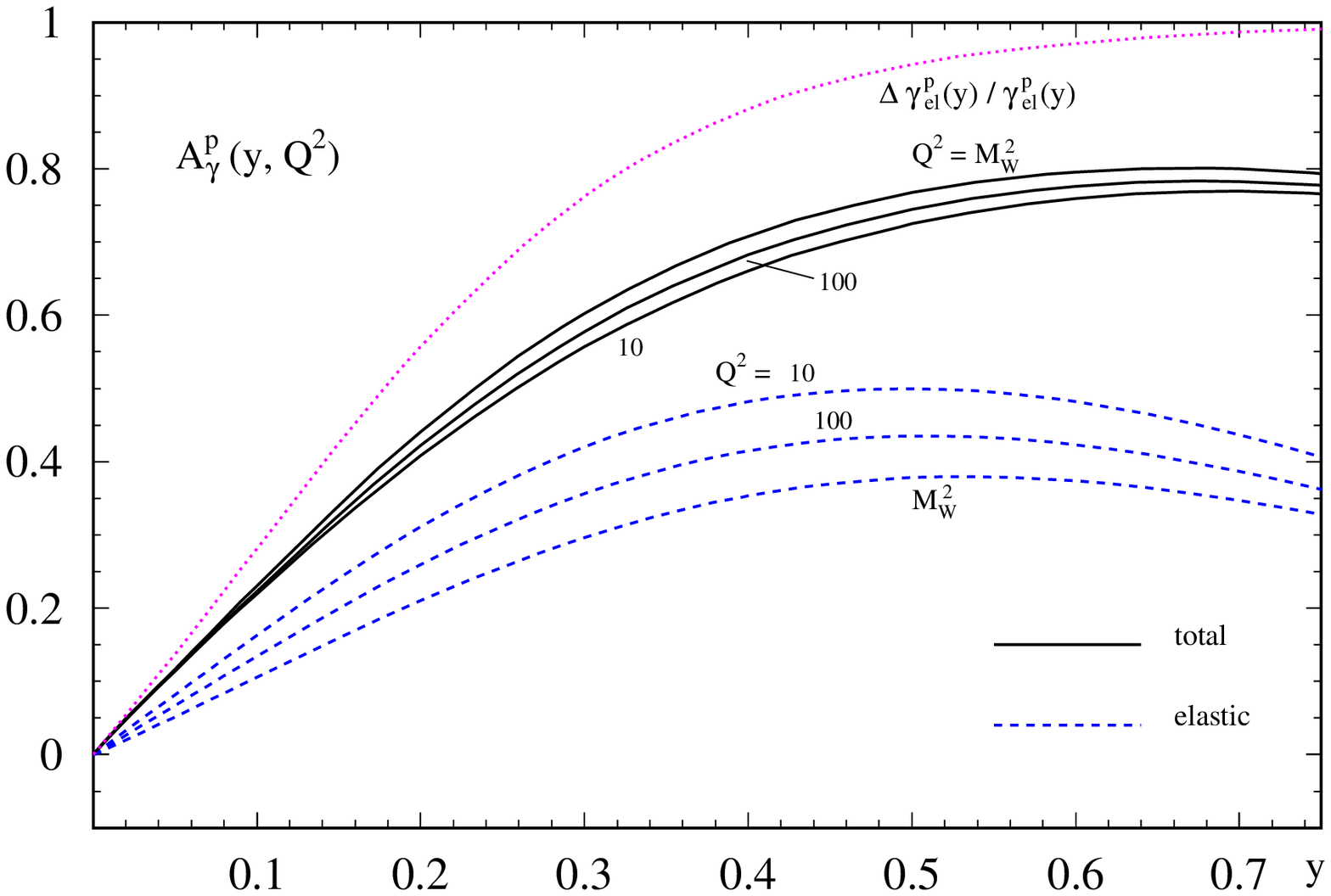,width=18cm} 
 
\vspace*{2cm} 
{\large\bf Fig. 3} 
\end{figure} 
 
\newpage 
\begin{figure}[t] 
\centering 
\hspace*{-1.5cm} 
\epsfig{figure=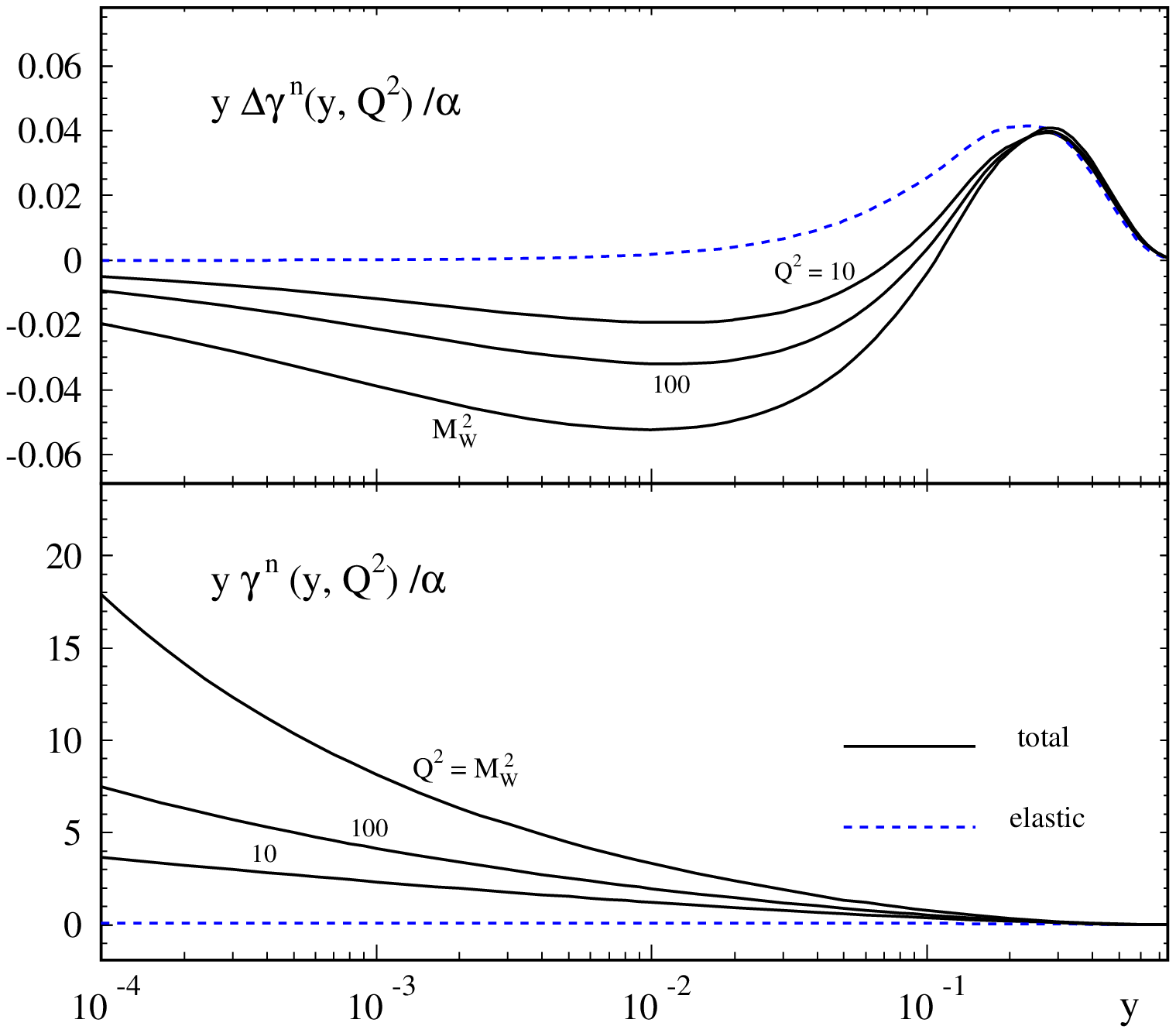, width = 18cm} 
 
\vspace*{2cm} 
{\large\bf Fig. 4} 
\end{figure} 
 
\newpage 
\begin{figure} [t] 
\centering 
\hspace*{-1.5cm} 
\epsfig{figure = 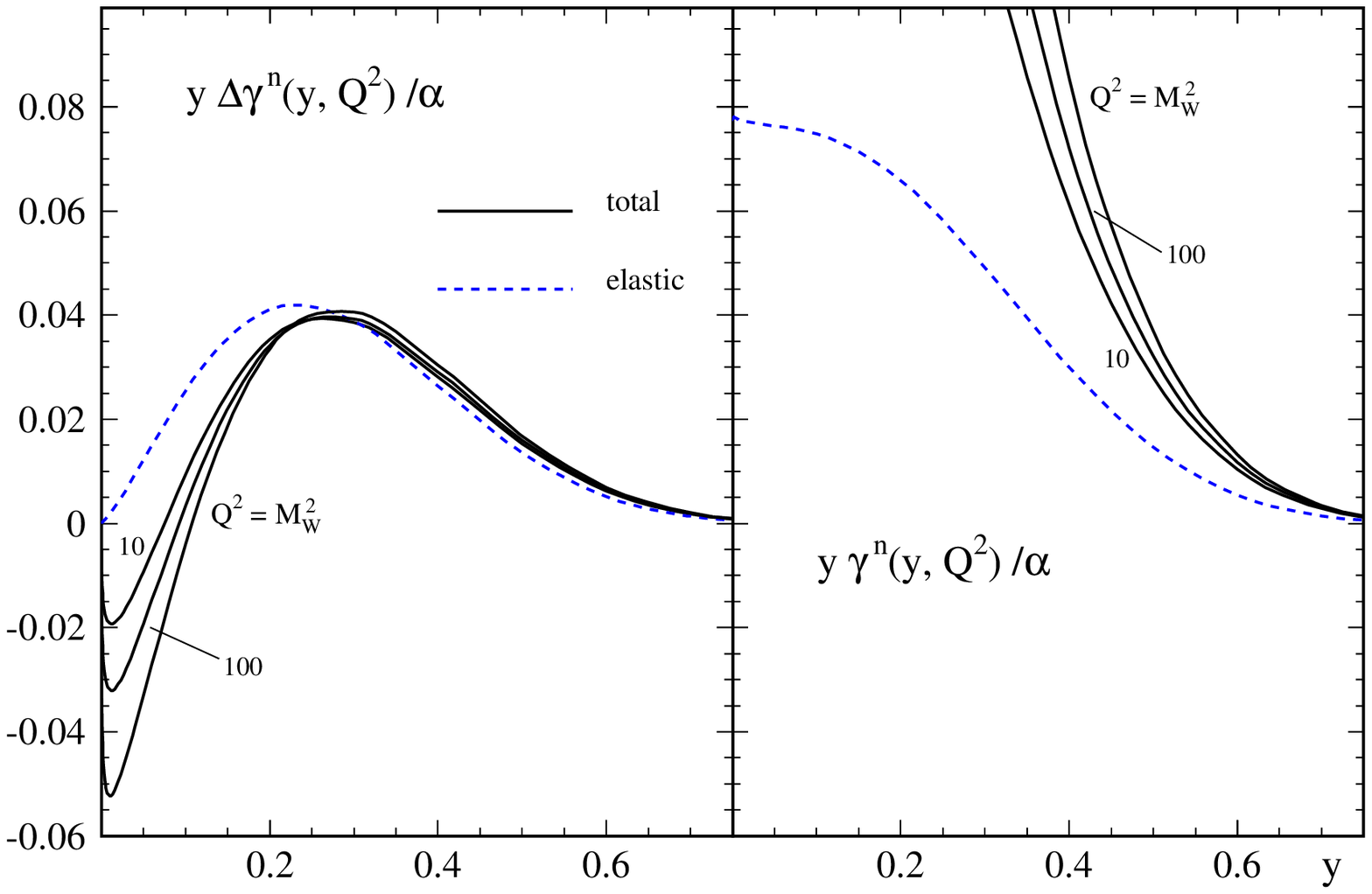, width = 18cm} 
 
\vspace*{2cm} 
{\large\bf Fig. 5} 
\end{figure} 
 
\newpage 
\begin{figure} [t] 
\centering 
\hspace*{-1.5cm} 
\epsfig{figure= 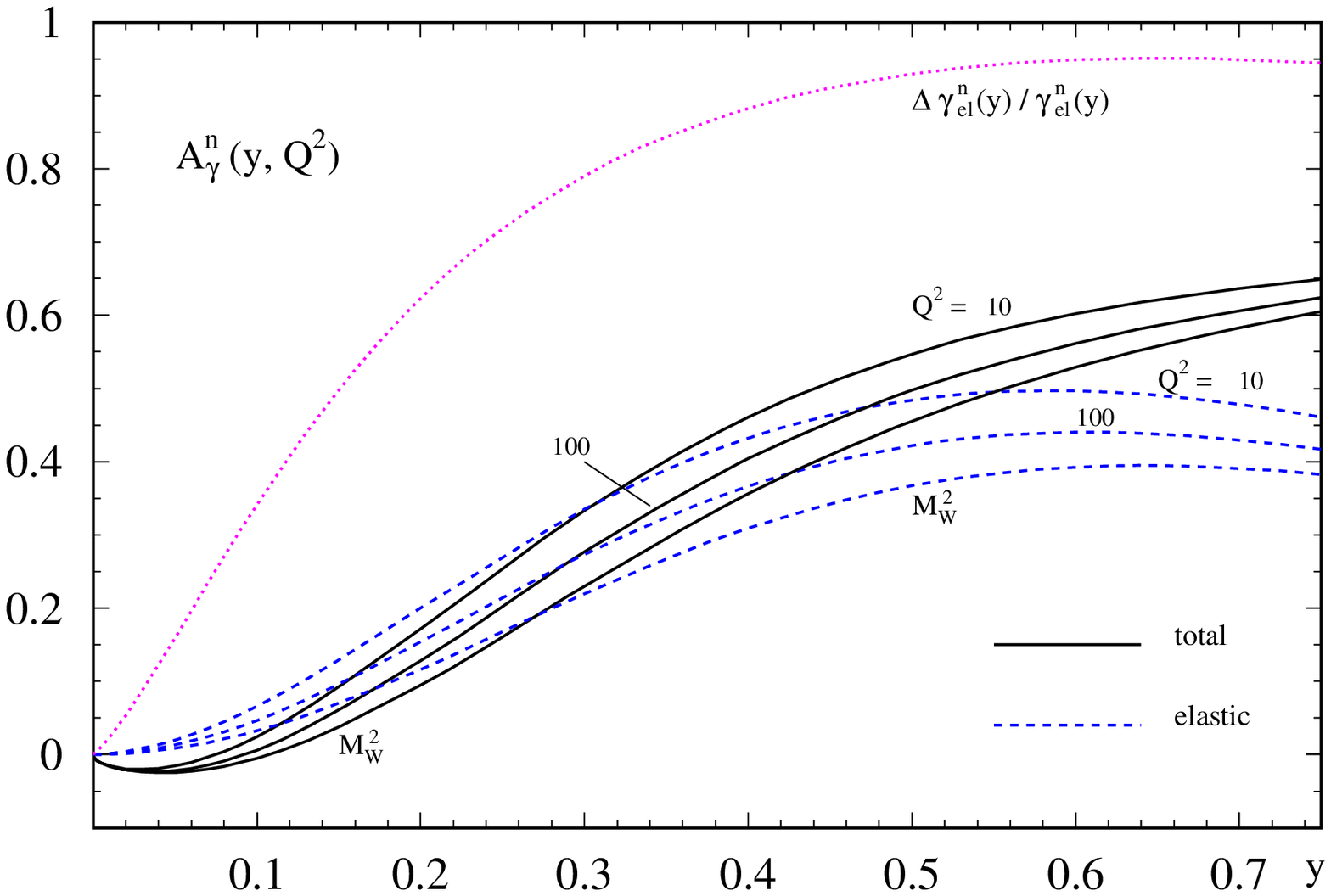, width = 18cm} 
 
\vspace*{2cm} 
{\large\bf Fig. 6} 
\end{figure} 

\newpage 
 
\begin{thebibliography}{54} 
\bibitem{ref1} C.F.\ Weizs\"acker, {\it Z.\ Phys.} 
        {\bf 88} (1934) 612;\\  
        E.J.\ Williams, {\it Phys.\ Rev.} {\bf 45} 
        (1934) 729 (L). 
\bibitem{ref2} B.A.\ Kniehl, {\it Phys.\ Lett.} 
        {\bf B254} (1991) 267.  
\bibitem{ref3} M.\ Gl\"uck, M.\ Stratmann, and 
        W.\ Vogelsang, {\it Phys.\ Lett.}  
        {\bf B343} (1995) 399. 
\bibitem{ref4} M.\ Gl\"uck, E.\ Reya, and A.\ Vogt,  
        {\it Eur.\ Phys. J.}  
        {\bf C5} (1998) 461. 
\bibitem{ref5} A.\ De Rujula and W.\ Vogelsang,  
        {\it Phys.\ Lett.} {\bf B451} (1999) 437.  
\bibitem{ref6} M.\ Gl\"uck, E. Reya, M.\ Stratmann,  
        and W.\ Vogelsang, 
        {\it Phys.\ Rev.} {\bf D63} (2001) 094005. 
\bibitem{ref7} D.\ de Florian and S.\ Frixione,  
        {\it Phys.\ Lett.}  
        {\bf B457} (1999) 236. 
\bibitem{ref8} S.\ Frixione, M.L.\ Mangano, P.\ Nason, 
        and G.\ Ridolfi, {\it Phys.\ Lett.} 
        {\bf B319} (1993) 339. 
\bibitem{ref9} M.\ Drees and D.\ Zeppenfeld,  
        {\it Phys.\ Rev.} {\bf D39} (1989) 2536.  
\bibitem{ref10} J.\ Bl\"umlein, G.\ Levman, and 
        H.\ Spiesberger,    
        {\it J.\ Phys.} {\bf G19} (1993) 1695. 
\bibitem{ref11} M.\ Drees, R.M.\ Godbole, M.\ Nowakowski 
        and S.D.\ Rindani, {\it Phys.\ Rev.} {\bf D50}  
        (1994) 2335;\\ 
        J.\ Ohnemus, T.F.\ Walsh and P.M.\ Zerwas, 
        {\it Phys.\ Lett.} {\bf B328} (1994) 369. 
\bibitem{ref12} C.E.\ Carlson and K.E.\ Lassila, 
        {\it Phys.\ Lett.} {\bf 97B} (1980) 291.  
\end{thebibliography}
\end{document}